# The Physics of Learning: From Autoencoders to Truly Autonomous Learning Machines

*Alex Ushveridze*

*Algostream Consulting, MN, USA*

alexush@algostream.com

## Abstract

*The fact that accurately predicted information can serve as an energy source paves the way for new approaches to autonomous learning. The energy derived from a sequence of successful predictions can be recycled as an immediate incentive and resource, driving the enhancement of predictive capabilities in AI agents. We propose that, through a series of straightforward meta-architectural adjustments, any unsupervised learning apparatus could achieve complete independence from external energy sources, evolving into a self-sustaining physical system with a strong intrinsic 'drive' for continual learning. This concept, while still purely theoretical, is exemplified through the autoencoder, a quintessential model for unsupervised efficient coding. We use this model to demonstrate how progressive paradigm shifts can profoundly alter our comprehension of learning and intelligence. By reconceptualizing learning as an energy-seeking process, we highlight the potential for achieving true autonomy in learning systems, thereby bridging the gap between algorithmic concepts and physical models of intelligence.*

## 0. Introduction

In recent years, machine learning (ML) has made significant strides toward creating systems that are not only smarter but also more self-reliant [1]. The most notable advancements in this area have largely stemmed from employing techniques such as meta-learning [2-4], self-supervised learning [5-7], reinforcement learning [8-10], and continuous learning [11-13]. The systems developed using these methods have been designed to enhance their own learning processes actively. They are equipped to function independently, adapt to new problems, and continuously improve themselves. For the sake of clarity, let's refer to these as *artificial autonomous learning* systems.

Alongside these developments, there's another dimension of autonomy that hasn't been as extensively explored in AI and ML discussions. This involves the ongoing inquiry into how the so-called *natural autonomous learning* systems might naturally arise from non-intelligent matter. While this question might appear theoretical at first glance, its potential practical implications are substantial. Deciphering the processes that lead to the spontaneous emergence of such systems and grasping the nature of the sustaining forces is crucial. This understanding not only expands our knowledge of the origins of intelligence in the universe but also opens up extensive practical opportunities for the creation and optimization of autonomous systems.

Methodologically, the two types of autonomy (artificial and natural) differ significantly. One source of this difference stems from the software-hardware dichotomy. Indeed, the design and development of artificial autonomous systems have largely focused on software, operating under the assumption that there exists a universal hardware platform capable of supporting a practically unlimited range of computations. Yet, for naturally evolving systems, the spontaneous generation of additional hardware resources, such as memory space or processing units for potential future needs, is highly unlikely (nature does not create things 'just in case'). The methodology used for exploring artificial autonomous learning systems overlooks the fact that, in nature, evolution simultaneously shapes both software (signal processing) and hardware (physical structures), allowing for adaptations beyond the capabilities of static hardware designs. The appropriate approach to this problem, in the context of natural autonomy, involves considering software-hardware cooperation.

Another source of distinction arises from the resource search and management problem. Note that all hardware modules require a steady, balanced, and reliable energy supply for proper functioning. However, in the artificial autonomous learning architectures, there is no specialized module designated solely to search for appropriate energy sources. The energy for all these modules is provided externally by humans in ways designed by humans in advance. Therefore, from the standpoint of any artificial autonomous device, the energy resource is taken for granted, meaning that its management problem is simply beyond the scope of artificial autonomous learning tasks. In contrast, for natural autonomous systems, the question of where to obtain energy and how to convert it into usable forms is central because it is directly linked to the system's functionality and survival.

The aim of this article is to initiate discussion on scientifically approaching both hardware-software cooperation and resource management problems. From a methodological standpoint, we believe it is more pragmatic to start with the latter—focusing first on the energy aspects of the learning process—simply because without correctly addressing this issue, speculations about the former would be baseless. In this article we will be primarily focused on just this energy aspect, and only briefly, in the last section, will touch the software-hardware cooperation problem and sketch a few possible ways of its solution within the energy-based approach we are going to discuss below.

The central idea guiding all subsequent logical constructions is that the task of finding resources in an unknown environment presupposes some rudimentary level of intelligence, encompassing essential capabilities such as learning (i.e., determining where energy might be found) and acting accordingly (i.e., extracting and utilizing the discovered energy). Moreover, the processes of learning and self-feeding cannot be temporally separated, as learning itself consumes energy and thus cannot occur prior to obtaining some form of nourishment. This presents a 'chicken-and-egg' logical loop. The resolution to this paradox lies in acknowledging a scenario where learning and energy consuming processes emerge simultaneously, as two inseparable aspects of a single phenomenon. With this conclusion as our starting point, we should explore what types of natural phenomena could embody this scenario.

It is evident that physics plays a crucial role in addressing this question. The endeavor to understand intelligence and learning — as an integral part of it — has historically been the domain of psychology, neuroscience, and artificial intelligence (AI). However, the undeniable truth is that intelligence is a phenomenon of the physical world and, as such, must adhere to physical laws and be studied within their framework. The efforts to connect intelligence with physics are not new, leading to the concept of self-organization, which is historically linked with chemical reactions and other complex dynamics of multi-particle systems. This phenomenon is typically examined through the lens of non-equilibrium thermodynamics — a theory that, because of its complexity and conceptual incompleteness, is still in the process of development [14]. This situation forges a strong association between intelligence and complexity. This link is further reinforced by the observation that the most advanced intelligent systems today, whether artificial or natural, are implemented within highly complex architectures.

All this has shaped a common belief currently dominating in the AI communities that intelligence is an extremely complex phenomenon and thus cannot be studied within simple systems. But what if that belief stems simply from our inability to grasp the fundamentals of intelligence? Beyond any doubt, intelligence is one of the most fundamental phenomena in the world. But if so, the physics underlying it must be simple – as fundamentality is almost a synonym for simplicity. Even more – for each fundamental phenomena one can always construct a certain simplest model in which this phenomenon is realized in purest possible form and can be studied within elementary methods. A good example is a hydrogen atom in quantum physics. What could play the role of such 'hydrogen atom' for the phenomenon of autonomous learning?

One of the possible candidates for such a role is a model grounded in the formalisms of two theories: thermodynamics of computation [15-18] and reversible computing [19-21], the two closely related branches of information theory focused on the study of energy expenditure and accumulation processes accompanying the process of computation. One of the pivotal outcomes of the first theory is that information, when accurately predicted, can be converted into energy. This discovery offers a compelling explanation for why an autonomous system might be inherently motivated to learn: by enhancing its predictive abilities, it can potentially increase its energy reserves for internal use [22,23]. However, predicting outcomes necessitates computations, which could incur additional energy costs. Is the energy gained through prediction enough to offset these costs? According to the outcome of the second branch, any computation can be executed in an energy-neutral manner, meaning that making predictions within this model does not result in any net energy expenditure [22,23], only in its gains. Owing to the distinct digital nature of the above considerations, we refer to this framework as *the digital autonomous model* in this paper.

Another (and, in our opinion, much more promising and interesting) candidate, embodies the concept of learning in an astonishingly simple and elegant manner. It is derived from classical mechanics and is realized in the form of simple resonant models with dissipation. These models can fully be described by low-dimensional systems of Newtonian equations of motion. Remarkably, this seemingly simple language suffices to describe the core effects of the learning process [23, 24]. Within these models, autonomous learning manifests as a process of energy extraction from the external environment through the process of resonance, without the necessity for digital mechanisms. We designate this framework as the *analog autonomous model* in this paper.

In this article, we venture to hypothesize that any unsupervised learning system, built on the conventional principles of AI and ML, can be transformed into a purely physical (analog) autonomous learning system. While we lack a general proof of this assertion, we demonstrate its plausibility by detailing the transformation of a well-known learning apparatus—the classical autoencoder [25-27]—into a physical model. This transformation unfolds in three distinct and meticulously planned steps, with minimal alterations at each stage to preserve the clarity of the logical progression and facilitate comprehension. This gradual approach underscores how incremental paradigm shifts can profoundly deepen our understanding of learning, a crucial aspect of intelligent behavior.

The paper is structured as follows: Section 1 examines the high-level architecture of the classical autoencoder, highlighting the challenges associated with its training methodologies, which, due to their external character, seem to be incompatible with the basic principles of autonomous learning. In Section 2, we present a modified version of the autoencoder, dubbed the 'parroting autoencoder', for its ability to replicate one of the most rudimentary learning methods: parroting. This altered architecture demonstrates considerable promise for internalizing the training, introduces the concept of internal learning loop – a key structural element on the road towards full energy autonomy, and also suggests avenues for investigating the interplay between subjective and objective perceptions of the external world. The latter topic, which as we think warrants further investigation, is briefly mentioned in Appendix. Section 3 transitions to the digital resonance model, which being an example of self-feeding autonomous learning model built on the base of internal learning loop, marks a significant shift from traditional ML methods. Its connection to conventional ML is kept primarily through its digital nature. Section 4 proposes a further departure from traditional ML paradigms by completely discarding the digital framework, arguing that it serves as an unnecessary complication to the inherent learning processes. In this section we discuss the 'analog resonance model' which can be realized within purely physical systems of resonant type. Finally, Section 5 delves into what we consider the heart of our inquiry: identifying the essential traits of a purely physical system that exhibits fully autonomous learning.

# 1. The Classical Autoencoder

Intelligence can be informally characterized by its ability to forge a reciprocal link between big data and small data. Here, "big data" refers to the observable attributes of real-world processes, while "small data" denotes our internal understanding (or, in other words, models) of these processes. This dual-pathway connection embodies two fundamental facets of intelligence: the capacity to learn, demonstrated by transforming extensive, real-world data into concise, internal knowledge, and the ability to apply this acquired knowledge practically, evident in the process of translating compact data back into actions or decisions that impact the larger world.

This concept is exemplified in one of the most significant breakthroughs in data science—the autoencoder [25-27]. An autoencoder (or, more exactly, its classical version, which we are planning to discuss in this section) is a specialized form of artificial neural network designed for the unsupervised learning of efficient data encoding. Its high-level architectural form is shown in the picture below:

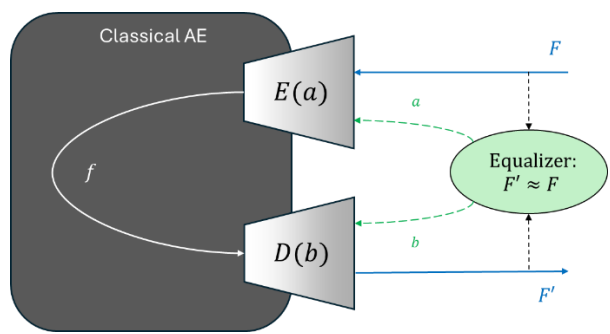

As it is seen, the classical autoencoder takes some external (big) data $F$ and converts it into its internal (small) representation $f = E(a)F$ utilizing a lossy compression algorithm $E(a)$ characterized by a set of tunable parameters $a$. This process aims to distill essential regularities from the original data $F$ thereby reducing the size of the resulting data $f$. This compression marks the completion of the first phase. The second phase involves decoding the compressed data $f$ by applying to it the 'inverse' transform $F' = D(b)f$. Here $D(b)$ signifies a decoder or decompression algorithm, also defined by (another) set of tunable parameters $b$. Hence, the transformation from the input $F$ to the output $F'$ can be conceptualized as the operation $F' = D(b)E(a)F$, with the overarching goal being to reconstruct the input data $F$ as accurately as possible through the output $F'$, striving for $F' \approx F$. Throughout the training process, the autoencoder autonomously learns to compress (encode) the input data and subsequently reconstruct (decode) it, aiming for a reconstruction that mirrors the original input as closely as possible. The model is fine-tuned to minimize the discrepancy $d(F, F')$ between the input and output, typically quantified using a loss function like the mean squared error. Minimizing this discrepancy is crucial for identifying the optimal parameters $a$ and $b$, ultimately yielding the most effective representation of the initial data $F$ through the reconstructed data $F'$.

One of the appealing aspects of autoencoders for data scientists lies in their versatility and implementation independence. Although autoencoders are predominantly utilized within neural networks, the specifics of their architecture are not crucial to their core concept. The construction of the encoding and decoding modules, along with the training algorithms used, are all secondary considerations. The primary focus is on the principle of creating a bottleneck between the input and output while ensuring the output remains similar to the input, despite this constraint. This bottleneck forces the system to autonomously decide what information is essential to pass through and what can be discarded to minimize loss. This simplicity and effectiveness underline the core appeal of autoencoders.

Despite the many benefits associated with autoencoders, they are not without significant limitations. A primary concern is their inherently linear structure (input – encoder – bottleneck – decoder – output), which necessitates external intervention for training. The mechanism that measures discrepancies between the input and output, crucial for training, operates externally, relying on data outside the autoencoder. This structure inherently limits autoencoders' capability to foster self-sustaining or internally motivated intelligence. For intelligence to be considered self-directing, the learning process must be governed by the agent's own subjective experiences. This would require the error measurement or any controlling function to be integrated within the autoencoder itself, rather than existing externally.

Following this line of thought, another challenge emerges concerning the determination of the error function to optimize the learning process's effectiveness. Regardless of how we measure the distance $d(F, F')$ between the input and output data, defining this distance using external data inherently limits its relevance and efficacy as learning progresses. For instance, starting with learning based on pixel comparison in images is perfectly valid. Yet, as learning advances to understanding basic shapes and their relationships, relying solely on pixel-level distances becomes inadequate. Ideally, the error function would adapt and evolve alongside the discovery of new data features. Unfortunately, the traditional autoencoder architecture, as previously illustrated, does not readily accommodate such dynamic adaptation of the error function[1].

---

[1] This issue can be somewhat mitigated through the use of "stacked autoencoders," a strategy employed in deep learning. Stacked autoencoders are essentially autoencoders arranged in layers or sequences, where the output from one layer serves as the input for the next, thereby creating a hierarchical structure of feature representations. These representations become progressively more abstract and compressed as data moves through the layers. A

Another challenge with autoencoders lies in their inherent need for a secondary cost function, beyond the basic error function $d(F, F')$. Solely focusing on minimizing the error function can inadvertently lead the autoencoder to learn the identity function, where data passes through the encoder and decoder unchanged, resulting in the internal representation mirroring the external data exactly. This outcome defeats the purpose of compression, indicating that effective learning requires not just any ransformation, but one that significantly compresses the data. The greater the compression, the more efficient and potentially insightful the learning process is deemed to be.

In the forthcoming section, we propose a reimagined approach to the autoencoder concept, transforming it into a tool conducive to intrinsic learning. We will address the three main challenges outlined: the dependence on external training, the static nature of the error function, and the necessity of additional cost considerations for compression. These issues can be resolved with modifications that, while substantial, do not radically alter the autoencoder's overall architecture. Crucial to these modifications is a reevaluation of the decoder's role and the complete internalization of the error control mechanism, thereby eliminating its external dependency.

## 2. The Parroting Autoencoder

The redesigned architecture of the autoencoder is depicted in the updated illustration below:

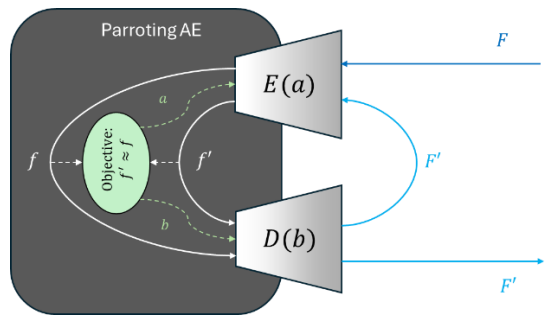

This new setup introduces a feedback loop that links the output of the decoder directly back to the input of the encoder, effectively internalizing the process that was previously governed by an external error function. This loop, which we hereafter will call the 'internal learning loop', will play a decisive role in framing up the true physical mechanism of learning. As we will see later, in the next two sections, this loop will evolve into an energy extraction engine ensuring a full autonomy of the learning device. Here we are still treating it in ML manner, as an internal error control and minimization mechanism. The fact that it operates solely with compressed data ensures that all adjustments and learning processes are based on the system's own, internalized criteria. This approach eliminates the need for external error evaluation, fostering a more autonomous learning environment within the autoencoder.

---

prevalent method for training stacked autoencoders involves greedy layer-wise training, which entails training each layer individually before integrating them. Once all layers have been pre-trained in this manner, the entire network undergoes fine-tuning to minimize the total reconstruction error. Nonetheless, even though this method enhances feature extraction and representation, it still relies on external training mechanisms. Consequently, it falls short of achieving the goal of autonomous and self-driven learning, as it does not fundamentally change the autoencoder's reliance on externally defined error functions or training processes.

To illustrate how this updated framework operates, let's conceptualize the encoder $E(a)$ and decoder $D(b)$ within the autoencoder as analogous to perception and communication mechanisms, akin to ears and mouth, respectively. These components, characterized by adjustable parameters $a$ and $b$, facilitate the autoencoder's interaction with external stimuli. Imagine the autoencoder encounters an external signal $F$, which is then processed by the encoder $E(a)$ to become the autoencoder's internal representation $f = E(a)F$ of that signal. If $F$ is an acoustic signal, for instance, then $f$ represents the sound as perceived internally by the agent. The characteristics of this perceived sound are determined by the encoder's parameters, making $f$ a subjective and private experience, inaccessible to external observation. In philosophical terms, $f$ is considered the agent's quale of $F$, highlighting the deeply personal nature of perception within this system.

Now, it is a good time to explain why we called the above schema the 'parroting autoencoder'. Parroting is one of the most fundamental mechanisms of learning, widely observed from human behavior to natural phenomena. It's a method heavily relied upon by children to match the sounds they make with those they hear, facilitating language acquisition and cultural assimilation. This mimicry isn't limited to human activities; it's a principle that extends to the learning of new disciplines, where individuals emulate the reasoning patterns of their instructors or the authors of their study materials. Beyond human behavior, nature itself employs replication extensively, not just in the biological replication of organisms or the formation of social groups but also through processes like resonance, spontaneous synchronization, and various forms of attractive forces. This universal tendency to imitate and replicate is a testament to the pervasive role of parroting as a learning strategy across different contexts and disciplines.

How do these concepts of mimicry and learning manifest within the described architecture? Consider an agent exposed to an external sound $X$. This sound once internally perceived as $x$, prompts the agent's vocal mechanism $D(b)$ to attempt reproducing the sound it just perceived. The outcome is a new external acoustic signal, represented as $F' = D(b)f = D(b)E(a)F$, paralleling the classical scenario. Here, both parameters $a$ and $b$ influence the process, yet $b$ is the uniquely external, adjustable parameter that the agent can deliberately modify to alter the produced signal's characteristics. The critical development here is that the autoencoder, or the agent, listens to the sound it has produced, denoted as $f' = E(a)F' = E(a)D(b)f = E(a)D(b)E(a)F$. This scenario, where the agent hears its own output, introduces two internal representations of the original sound: $f$ and $f'$. While these may not initially align, the agent's goal is to converge these perceptions as closely as possible, effectively mimicking the original signal. This aim of aligning $f$ with $f'$ forms the basis for optimizing parameter $b$, a process fully within the agent's control based on its internal, subjective experiences. The optimization criterion, minimizing the distance $d(f', f)$ between the two internal representations, mirrors the error minimization approach seen in traditional autoencoders. However, the significant distinction lies in the source and application of the error function $d(f', f)$, which now revolves around the agent's subjective experiences—the comparison of its internal qualia $f'$ and $f$.

Let's delve into the optimization process. It's crucial to understand that the parameters $a$ and $b$, which define the encoder and decoder, are subject to distinct optimization strategies. Specifically, optimizing $a$ enhances the encoder's ability to recognize patterns within the external signal, leading to more efficient compression. This optimization process for $a$ is driven by a unique goal: maximizing the signal compression rate, a metric solely dependent on the properties of the external signal $F$ and unrelated to parameter $b$. Consequently, the improvement of the encoder (hearing abilities) and the decoder (reproduction abilities) are fundamentally different objectives, necessitating separate optimization procedures with their respective objective functions. This distinction underlines the conceptual separation between training the encoder and decoder, each with its tailored approach to optimization.

The concept of parroting, at first glance, appears straightforward, intuitive, and full of potential, especially when considered from a broad, meta-architectural viewpoint. From the purely philosophical standpoint, the parroting autoencoder provides an easy way of explaining the appearance of dual objective/subjective images of external objects as a result of communication between multiple parroting devices. For more detail on the corresponding mechanisms see Appendix.

Note that within the conventional realms of machine learning and artificial intelligence, implementing the concept of parroting autoencoder seems feasible without facing any overwhelming challenges. However, a deeper examination, especially in light of our initial goal to transition beyond the classic autoencoder model towards a parroting variant, reveals a complex landscape. Our aim was to unearth an architecture that supports self-learning, granting AI agents the capability to independently commence and maintain their learning journey. A critical requirement for such autonomy is the device's complete energy independence from any external user (here, "user" refers to any entity outside the autoencoder that supplies energy for its operation towards achieving specific tasks).

Yet, even a cursory examination of the autoencoder's structure, whether in its classical or parroting form, indicates not just an energy reliance but a conceptual energy dependence. This dependency fundamentally clashes with the notion of energetic autonomy, as dictated by foundational principles.

Surprisingly, this challenge stems directly from a pivotal component of the autoencoder's operation: its encoder. This module, tasked with performing lossy compression of the input signal, embodies the core issue. Lossy compression, by its nature, is a logically irreversible process. However, research in the thermodynamics of computation has shown that logical irreversibility leads to thermodynamic irreversibility, which inevitably results in the dissipation of energy that cannot be recovered. The energy loss incurred during compression can be quantified as $E = kTN \ln 2$, where $N$ represents the number of bits of information lost. This relationship between lossy compression and energy loss underscores a fundamental obstacle to achieving energy independence within the autoencoder framework.

This realization brings us to the following conclusion: correctly addressing the above issue and thus unblocking the road towards practical realization of autonomous learning concept cannot be achieved through algorithmic solutions alone. Incorporating physical, specifically energy-based, considerations is essential. One of the foundational theories that can at least partially tackle such challenges is "computational thermodynamics," which assesses the energy efficiency of computational processes. This theory provides measurable energy costs for each basic logical operation and read-write action. In the following section, we will apply the principles of computational thermodynamics to reconfigure the architecture of the parroting autoencoder, aiming to enhance its energy efficiency.

## 3. Learning as Digital Resonance

The challenge of autonomous learning that we're exploring in this article is fundamentally rooted in the field of physics, specifically within the branch known as non-equilibrium thermodynamics. This area of study aims to uncover the processes behind self-organization in open physical systems, aspiring to elucidate a wide range of phenomena. This includes the origins and development of life and the emergence and evolution of intelligence from inanimate and non-intelligent matter. It's important to acknowledge, however, that the lofty ambition of fully understanding these complex processes remains a distant goal, far from being achieved at present.

Despite the current lack of a complete and self-consistent theory in non-equilibrium thermodynamics, it doesn't mean we should avoid exploring preliminary models or "proxies" that draw upon concepts from established physical theories. These proxies, while they might not achieve full internal coherence, provide valuable early insights into the mechanisms that could enable non-intelligent systems to exhibit behaviors akin to intelligence. Computational thermodynamics stands out as one such proxy, born from the fusion of information theory and classical thermodynamics. Its key propositions include:

- *Information possesses physical properties and can act as an energy source when its outcomes are predictable. Each bit of accurately predicted information can be transformed into a quantifiable amount of usable energy, specifically $E = kT \ln 2$.*
- *The deletion of information (understood as the 'reset' operation) incurs energy losses, with the energy cost of erasing each bit of information mirroring the above equation: $E = kT \ln 2$.*
- *Creating a new a-priori known information 'from scratch' – in a memory cell with unknown content – requires the same amount of energy, $E = kT \ln 2$, per bit.*
- *The inverse process of randomization of information stored in a memory unit does not require any energy: it is completely energy neutral.*
- *Any computational process that is logically reversible, meaning it doesn't result in any net change in information quantity, can be conducted in an energy-neutral manner.*

Strictly speaking, these principles are valid if the corresponding operations on memory units are performed adiabatically slowly. However, even so, they underline the significance of grasping and leveraging the physical aspects of information as a cornerstone for creating systems capable of autonomous learning and operation, with the added benefit of energy efficiency or neutrality. They elucidate the intrinsic motivation for an agent to engage in learning: it presents an opportunity to transform received information—specifically, its predictable components—into energy. This conversion unlocks a universal resource that the agent can exploit to fulfill any conceivable need that can be algorithmically defined.

A critical takeaway from these concepts is the irrelevance of the information's content in contrast to the agent's capability to accurately predict it. This distinction is pivotal because it directly ties an agent's motivation to learn with its quest to improve its predictive accuracy, thereby granting it access to previously untapped energy reserves. Moreover, these principles advocate for actions, messages, and statements to be based on pre-existing information, embodying what is commonly understood as rational thinking. By doing so, the agent avoids generating new information, circumventing the associated energy costs and losses. This strategy ensures that learning and decision-making processes not only conserve energy but are also directed towards maximizing the agent's energy acquisition and utilization efficiency.

The insights derived from these principles lead to the confirmation of our earlier statement: both the classical and the parroting versions of the autoencoder are markedly inefficient in terms of energy consumption. At first glance, this realization might seem to undermine the feasibility of employing autoencoders, particularly those with a parroting mechanism, in self-sustaining learning devices.

Yet, it's important to ask ourselves: should we only consider the negative implications of these findings? Could these principles also offer a constructive pathway, guiding us out of this apparent deadlock?

Fortunately, the response is encouragingly affirmative. Moreover, a deeper exploration reveals that these principles do not merely provide means to mitigate energy losses; they pave the way for transforming the parroting process from an energy-draining activity into one that actively generates energy. This essentially converts autoencoders into autonomous learning systems, turning a critical weakness into a groundbreaking strength. To clearly illustrate the concept of transforming energy losses into energy gains, let's introduce a revised block-schema of the parroting autoencoder, diverging from the version discussed previously in Section 2.

The revised schema, which we're introducing below, bears a resemblance to its predecessor discussed in the previous section by mimicking the functionality of the identity operator, a simple yet profound concept. In this schema, the external signal $F$ emerges unchanged after traversing two computational stages, previously identified as $E(a)$ and $D(b)$. However, unlike their roles as encoder and decoder in the earlier model, in this context, they serve as reversible computing units tasked with performing subtraction operations in a reversible manner.

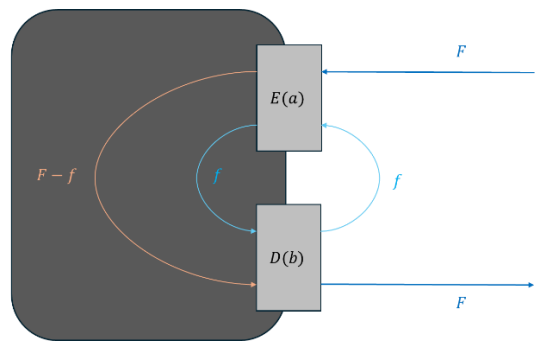

This setup ensures that the output from $E(a)$, which includes the difference $F - f$ between the two input signals $F$ and $f$, still retains the input signal $x$, thereby enabling the restoration of both initial inputs. The same principle applies to the subsequent block, $D(b)$. The key feature here is the reversibility of these computational blocks, rendering their operation energy-neutral. This approach not only conserves energy but also exemplifies how computational processes can be designed to align with principles of energy efficiency and reversibility:

Let's shift our perspective slightly and view these two blocks as components characterized by parameters $a$ and $b$, which regulate the strengths and character of the internal signal $x$ within the schema. This internal loop, with signal $f$ exiting block $D(b)$ and re-entering through $E(a)$, essentially mirrors the process described earlier. The primary objective of this setup is to refine the internal signal $f$ to closely mirror the external signal $F$, embodying an attempt at parroting with a new, resource-oriented motivation. In an ideal scenario, where the internal representation $x$ perfectly replicates the external signal $F$, their difference diminishes to zero, yielding a constant signal of paramount importance for energy generation. This scenario leverages the principle that a constant and thus fully predictable signal enables the extraction of maximum energy from each bit, according to Landauer's principle, amounting to $E = kT \ln 2$. By integrating an energy extraction unit within this process, as depicted in the updated schema:

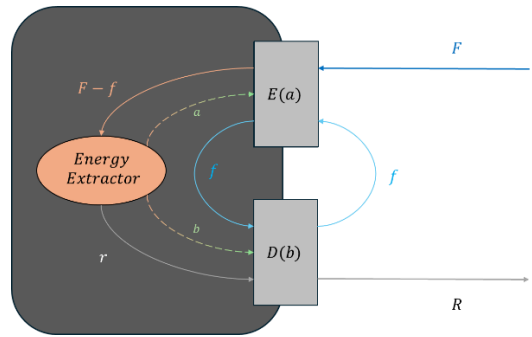

the autoencoder can directly translate its parroting capabilities into tangible energy flows. This energy, harvested from the act of internal data reconstruction (parroting), can be allocated towards any operational needs of the autoencoder, most notably towards refining its signal reproduction algorithms. This approach not only highlights a novel method of energy conservation but also underscores the potential for self-sustaining learning mechanisms.

It's important to recognize that the act of energy extraction introduces randomness to the input signal, effectively transforming it into an output with zero informational content. Consequently, incorporating an energy-extraction component effectively transitions the identity operator into a dissipative system. This transformation aligns perfectly with the second law of thermodynamics, which posits that the natural processes—including those of life, learning, and intelligence—do not merely comply with but actively promote the second law. This perspective underscores the fundamental role of energy dynamics in driving the phenomena of life and intelligence, situating these processes as natural facilitators of thermodynamic principles.

The phenomenon we've outlined above can be termed "digital resonance" for its clear digital nature (on one hand) and its resemblance to the classical resonance observed in physical systems (on the other hand). This learning process entails matching patterns from an external digital signal $F$ with those in an internally generated digital loop signal $f$, thereby enabling energy extraction in a way that mirrors resonance in physical systems. At the same time, this process unfolds within a standard computational framework, where all operations inherently possess a digital character.

Given this context, questioning the necessity and practical implications of these signals' 'digitality' and their processing becomes relevant. Does the digital nature of these signals facilitate more efficient energy extraction? Does it render the process more transparent? Closer examination reveals that the digital characteristics of these signals do not markedly impact the energy extraction process or enhance the system's overall efficiency. Furthermore, it becomes apparent that digitality is not an essential aspect of this phenomenon, leading us to question its relevance. This insight paves the way for a more profound proposition: the learning process can be understood and elucidated purely through the lens of physics, independent of information theory. This perspective advocates for a foundational shift towards a physical explanation of learning processes, moving beyond the specific focus on digitality. These are the questions we are going to discuss in the next section.

# 4. Learning as Analog Resonance

Before we start, let's first recall the definition of learning given in ref. [23], in which it was characterized as "a process wherein a system attempts to replicate the dynamic patterns of external forces using entirely internal mechanisms". We see that this definition essentially mirrors the concept of parroting discussed earlier, resonating with our intuitive understanding of learning as an internally motivated process. At the same time, by framing this process in terms associated with forces and their dynamics, this definition highlights the intrinsic link between learning and physics.

To better understand the nature of this link, it's crucial to recognize that any learning system, or 'learner,' operates as an open physical system in active interaction with its environment, which plays the role of its 'teacher.' This interaction is characterized by external forces applied by the teacher and the learner's responsive actions. During these exchanges, the energy within the learner is not necessarily conserved; it can either decrease due to destructive forces or increase through constructive engagement. Now, consider a learner aiming to augment its internal energy by strategically engaging with its surroundings. In theory, this could be achieved either by altering the environment or by the learner modifying its own internal structure. While large, powerful systems might have the luxury of choosing either approach, smaller, less powerful learners may only feasibly opt for self-adaptation to external forces. To address the question of necessary internal modifications for adaptation, physics offers a compelling answer: the learner can leverage the phenomenon of resonance, one of the most recognized effects in physics. The core principle of resonance is the synchronization of movements between two interacting systems, which dramatically amplifies the energy exchange between them. In this dynamic, the smaller system stands to gain substantially, as it can enhance its internal energy by efficiently absorbing energy from the larger system. This process of resonance not only facilitates a more significant energy transfer but also exemplifies how strategic alignment with external forces can lead to an advantageous increase in a system's internal energy reserves.

In educational contexts, resonance is typically discussed with reference to the impact of periodic external forces on harmonic (linear) oscillators, with its practical uses often confined to tasks involving frequency matching. Yet, the concept of resonance encompasses a far wider scope than these specific instances suggest. To explore this, consider the learner as a mechanical system interacting with its environment via a dynamic variable, labeled here as the position variable $x$. Classical mechanics tells us that the change in the system's internal energy over time, $dt$, is dictated by the mechanical work done by an external force $f$ acting on $x$ within that timeframe. This work is quantified by the equation: $E = f dx$, where $dx$ is the change in $x$ over $dt$. Dividing both sides of this equation by $dt$ yields $dE/dt = fv$, with $v$ representing the system's velocity. It's clear from this formulation that for the learner to augment its energy, the product of the external force and the system's velocity, represented on the right side of the equation, must remain positive. Given that both $f$ and $v$ vary over time, maintaining the positivity of this product necessitates that the system's velocity $v$ aligns or synchronizes with the external force $f$, meaning their behaviors should mirror each other. Symbolically, this can be denoted as: $v \sim f$.

Termed the 'resonance condition,' this relationship offers a straightforward strategy for smaller systems to harness energy from larger ones, such as the external environment. Achieving this requires the system to emulate the behavior of the external force $f$ through one of its internal variables, like velocity $v$, effectively adopting a resonance-based approach to energy extraction.

The discussion so far reveals that the resonance effect naturally fulfills the core objective of learning: replicating external processes through entirely internal mechanisms. This once enigmatic process now receives a straightforward physical explanation. It becomes evident why smaller systems might 'desire' to emulate the behavior of larger systems: such 'parroting' inherently leads to the accumulation of energy. This realization demystifies the motivation behind smaller systems' mimicry of larger ones, grounding it in the tangible benefit of energy gain.

Note that the parroting (the generalized copying process), can naturally be treated as one of the simplest forms of self-organization, because the adaptation to the regularities observed in the external environment increase the internal order of the learning system. According to the second law of thermodynamics, any increase of order in the system's internals must be accompanied by an increase in disorder in its surroundings. This means that the resonance-based model of learning cannot be complete without explicitly incorporating in it the process of dissipation. In digital version of learning we considered in previous section, we dealt with dissipation in the form of a no-information-bearing randomized output signal. This randomization was considered exclusively in thermodynamic context. Here, in the analog version of learning, based on the low-dimensional model of classical mechanics, there is no room for thermodynamics as such, however the effects of dissipation can nevertheless be correctly taken into account through local friction forces acting on system's degrees of freedom. The equation describing the energy change rates in the dissipative case reads $dE/dt = fv - \gamma v^2$, where $\gamma$ denotes the friction coefficient.

If the external force $f$ varies over time, the system can only increase its internal energy by starting to move – i.e. changing its velocity $v$ from zero to some non-zero values. However, not any changes of velocity would work – only those which are synchronized with the external force: $v \approx cf$ (with some positive coefficient $c$) and which are sufficiently small (so the first term in the right-hand side still dominates over the second one). If the increase of $v$ will continue, the energy eventually stops growing. This happens at the moment when the dynamical equilibrium of the input and output flows of energy will be established. Note that in this case, still characterized by the same condition $v \approx \gamma^{-1} f$ both these flows as well as the total energy accumulated in the system reach their maximum.

## 5. Discussion and the Next Steps

The core idea we have endeavored to convey in this article is the reconceptualization of learning as a form of resource-seeking behavior. Generally, the distinction between seeking knowledge and resources is largely semantic, as knowledge is often regarded as a resource. However, in our context, this distinction takes on a unique significance. By saying that the motivation of systems to learn can be explained through the motivation of having more resource, we do not mean a generic and typically vague concept of resource but rather its very tangible and quantifiable form: the energy. This perspective is critical, because it allows us to understand the dynamics of learning through dynamics describable within the laws of physics, providing an alternative framework for exploring the phenomenon of learning. This obviously complements traditional approaches in AI/ML, neuroscience, and psychology. Within this purely physical framework, the term 'motivation' obviously transcends its metaphorical usage, approaching a nearly physical concept that describes the forces driving open systems (learners) to interact with their environments (teachers) in manners akin to learning.

Translating the learning process from the domain of information to that of physics underscores its inherently dynamic character. A pivotal aspect of this dynamism and its driving force is that any successful learning act—such as the accurate internal reproduction of external patterns—is immediately rewarded with an influx of energy. The magnitude of this energy flow is directly proportional to the 'quality' of learning, thereby providing immediate positive feedback for any learning attempt. This mechanism could theoretically account for the emergence of 'self-learning centers' as stable yet dynamic fluctuations within the system, akin to twirls formed by learning loops. The perpetuity of these twirls is sustained by the presence of regularities in external forces acting upon the system. The temporal patterns of energy accumulation and dissipation within these twirls mirror the temporal profile of the external forces, illustrating a profound synchronicity between the system and its environment. And this synchrony is exactly what characterizes the process of learning in the traditional sense of this word.

The vision outlined offers compelling groundwork for developing a model that explains the spontaneous emergence of fully autonomous self-learning structures from non-living matter. Yet, there's an important nuance to consider. The described synchrony between internal and external data patterns, along with the equilibrium of energy flows, signifies a state of resonance. However, a question arises when considering systems initially distant from this state: What drives such a system towards resonance? How can a system foresee the fulfillment derived from attaining a resonant state, and what prompts it to modify its structure in pursuit of this goal?

These inquiries delve into the core of understanding how a system can inherently adjust itself to achieve resonance, a concept that spans both physical theories and machine learning paradigms. The concept of "tuning" is central in both domains, pointing towards the need for mechanisms that facilitate self-tuning within a system. Imagining the learning system as divided into two components—a resonator, where data matching and energy extraction transpire, and a controller, which adjusts the resonator's parameters to align with varying external conditions—helps clarify this idea. In the context of the software-hardware dichotomy, the activities within the resonator can be likened to software operations, whereas the resonator and its controller clearly constitute the hardware aspects. The challenge lies in fostering a seamless interaction between software and hardware, ensuring that both the resonator and the controller act as integrated parts of a single physical system, propelling it towards resonant states as defined by the external data's characteristics.

So, how do we achieve such a sophisticated level of integration and self-tuning? At present, several approaches appear promising, potentially offering solutions within the analog resonant learning model framework. These strategies aim to identify and harness physical mechanisms that can naturally lead a system towards achieving and maintaining resonance, thereby enabling a form of intrinsic learning and adaptation reflective of the dynamic interplay between a system and its environment. Let us briefly describe some of them.

Let us start with the concept of multi-resonant media providing an intriguing approach to self-tuning necessary for spontaneous learning in autonomous systems. The system based on this concept can be likened to an extensive Kirchhoff network featuring randomly distributed capacitances, inductances, and resistances. The quantity of independent loops in such a network grows approximately as a square of the number of junctions. This architecture results in a spectrum of resonant circuits, each tuned to its unique eigenfrequency, rendering the system highly responsive to a wide array of signal patterns (akin to a Fourier expansion of the learned signal). This enables the system to extract energy from a diverse range of external signals, showcasing the potential of multi-resonant media as a powerful model for learning and energy extraction through resonance. By adding random nonlinearities to the network may significantly extend the range of external signals it can resonate with.

Another interesting concept – the noise-driven tuners – is based on the idea of leveraging phenomena traditionally viewed as detrimental—random fluctuations and braking systems—to facilitate constructive outcomes. Typically, these elements are associated with chaos and dissipation, concepts seemingly antithetical to order and self-organization. However, when harnessed together, random fluctuations and braking mechanisms can theoretically serve a constructive purpose by aiding in the search for parameter configurations that yield higher energy outputs, such as resonant states. Indeed, imagine a scenario where random fluctuations prompt the tunable parameters of a system to explore the parameter space through a stochastic process, akin to a random walk. The braking system then acts as a regulatory mechanism, arresting this random exploration upon the system's entry into a resonant state, which is characterized by increased energy production. This process suggests a symbiotic relationship between randomness and control, where random fluctuations expand the system's exploratory range, and the brake system ensures stabilization upon reaching energetically favorable configurations.

The last concept we want to briefly describe here exploits the idea of pattern-to-pattern attraction. It introduces a fascinating theoretical speculation that pushes the boundaries of how we might conceive of autonomous learning processes in physical systems. While this idea currently resides more in the realm of theoretical exploration, its potential evolution into a more tangible concept with further development should not be dismissed. Envision a hypothetical scenario where similar patterns inherently attract each other. This could be analogized through the physical construct of an absolutely elastic thread composed of magnetic dipoles. This thread serves as an information carrier, where sequences of dipoles with north (N) and south (S) orientations represent binary data (0s and 1s, respectively). The intrinsic attraction between like orientations (N-N and S-S) would naturally draw similar sequences closer together due to the pattern-to-pattern attractive forces. Now imagine one end of this thread is left free. It would spontaneously begin to coil into a ball, driven by the aggregation of similar dipole sequences. This growing ball, continuously collecting longer sequences, could metaphorically represent a learner. The process of this ball winding along the thread, expanding as it assimilates more like patterns, could be interpreted as an autonomous learning process. This conceptual model suggests that learning could proceed autonomously, driven solely by the intrinsic forces generated by the informational content encoded on the thread.

The commitment to delve into the three aforementioned concepts in future publications sets an ambitious path toward advancing our comprehension of autonomous systems. By exploring these innovative avenues, the objective is to elucidate the conditions under which systems can self-organize and achieve genuine functional autonomy. This endeavor is not merely an expansion of our knowledge regarding learning and energy harnessing in physical systems; it is a call to fundamentally rethink the underlying principles of intelligence as it manifests throughout the universe.

## References

[1] Yann LeCun, "A Path Towards Autonomous Machine Intelligence, Version 0.9.2", Openreview 2022-06-27, https://openreview.net/pdf?id=BZ5a1r-kVsf
[2] Timothy Hospedales, Antreas Antoniou, Paul Micaelli, Amos Storkey, "Meta-Learning in Neural Networks: A Survey", arXiv:2004.05439 [cs.LG], (2020)
[3] Nathan Hu, Eric Mitchell, Christopher D. Manning, Chelsea Finn, "Meta-Learning Online Adaptation of Language Models", arXiv:2305.15076 [cs.CL], (2023)

[4] Louis Kirsch, James Harrison, Jascha Sohl-Dickstein, Luke Metz, "General-Purpose In-Context Learning by Meta-Learning Transformers", arXiv:2212.04458 [cs.LG], (2022)
[5] Jie Gui, Tuo Chen, Jing Zhang, Qiong Cao, Zhenan Sun, Hao Luo, Dacheng Tao, "A Survey on Self-supervised Learning: Algorithms, Applications, and Future Trends", arXiv:2301.05712 [cs.LG], (2023)
[6] Randall Balestriero, Mark Ibrahim, Vlad Sobal, Ari Morcos, Shashank Shekhar, Tom Goldstein, Florian Bordes, Adrien Bardes, Gregoire Mialon, Yuandong Tian, Avi Schwarzschild, Andrew Gordon Wilson, Jonas Geiping, Quentin Garrido, Pierre Fernandez, Amir Bar, Hamed Pirsiavash, Yann LeCun, Micah Goldblum, "A Cookbook of Self-Supervised Learning", arXiv:2304.12210 [cs.LG], (2023)
[7] Xiao Liu, Fanjin Zhang, Zhenyu Hou, Zhaoyu Wang, Li Mian, Jing Zhang, Jie Tang, "Self-supervised Learning: Generative or Contrastive", arXiv:2006.08218 [cs.LG], (2020)
[8] Sanjeevan Ahilan, "A Succinct Summary of Reinforcement Learning", arXiv:2301.01379 [cs.AI], (2023)
[9] Vincent Francois-Lavet, Peter Henderson, Riashat Islam, Marc G. Bellemare, Joelle Pineau, "An Introduction to Deep Reinforcement Learning", arXiv:1811.12560 [cs.LG], (2018)
[10] Hui Bai, Ran Cheng, Yaochu Jin, "Evolutionary Reinforcement Learning: A Survey", arXiv:2303.04150 [cs.NE], (2018)
[11] Liyuan Wang, Xingxing Zhang, Hang Su, Jun Zhu, "A Comprehensive Survey of Continual Learning: Theory, Method and Application", arXiv:2302.00487 [cs.LG], (2023)
[12] Matthias De Lange, Rahaf Aljundi, Marc Masana, Sarah Parisot, Xu Jia, Ales Leonardis, Gregory Slabaugh, Tinne Tuytelaars, "A continual learning survey: Defying forgetting in classification tasks", arXiv:1909.08383 [cs.CV], (2019)
[13] Md Yousuf Harun, Jhair Gallardo, Tyler L. Hayes, Christopher Kanan, "How Efficient Are Today's Continual Learning Algorithms?", arXiv:2303.18171 [cs.CV], (2023)
[14] Adam Rupe and James P. Crutchfield, "On Principles of Emergent Organization", arXiv:2311.13749v1 [cond-mat.stat-mech], (2023)
[15] L. Szilard, "Uber die entropieverminderung in einem thermodynamischen system bei eingriffen intelligenter wesen", Z. Phys. 53, 840 (1929).
[16] R. Landauer, "Irreversibility and heat generation in the computing process", IBM J. Res. Dev. 5, 183 (1961).
[17] Charles H. Bennett, "Thermodynamics of Computation -- a Review", International Journal of Theoretical Physics, Vol 21, No 12, 1982
[18] Plenio, M. B. & Vitelli, V., 2001. The physics of forgetting: Landauer's erasure principle and information theory. Contemporary Physics, 42(1), pp. 25-60.
[19] Fredkin, E. & Toffoli, T., 1982. "Conservative Logic". International Journal of Theoretical Physics, 21 (3-4), p. 219–253.
[20] Feynman, R. P. & Hey, A., 2000. "Feynman Lectures On Computation". Westview Press; Revised ed.
[21] Perumalla, K., 2014. "Introduction to Reversible Computing". Chapman & Hall/CRC Computational Science.
[22] Alex Ushveridze, "Can Turing machine be curious about its Turing test results? Three informal lectures on physics of intelligence", arXiv:1606.08109, (2016)
[23] Alex Ushveridze, "Understanding Learning through the Lens of Dynamical Invariants", arXiv:2401.10428 [cs.AI], (2024)
[24] Alex Ushveridze, "On Physical Origins of Learning", arXiv:2310.02375 [q-bio.NC], 2023
[25] Dor Bank, Noam Koenigstein, Raja Giryes, "Autoencoders", arXiv:2003.05991 [cs.LG] (2020)
[26] Umberto Michelucci, "An Introduction to Autoencoders", arXiv:2201.03898 [cs.LG], (2022)
[27] Gabriele Martino, Davide Moroni, Massimo Martinelli, "Are We Using Autoencoders in a Wrong Way?", arXiv:2309.01532 [cs.LG], (2023)

# Appendix. Collective Perception in Parroting Autoencoders

In this appendix, we explore a scenario where multiple parroting autoencoders communicate to 'discuss' their observations of an external object or event, $F$. Our focus is on the mechanism through which an objective image of this event, which we denote as $\text{Obj}(F)$, emerges from the subjective images generated by each autoencoder.

Consider a certain large set of autoencoder agents $A_i$, each equipped with an encoder-decoder pair

$$\{E_i, D_i\}, \quad i = 1, \ldots, N$$

and assume that all receive an external signal $F$. Each agent generates an internal visual image of $F$ represented by

$$f_i = E_i F, \quad i = 1, \ldots, N$$

These visual images are inherently private and subjective, thus not directly sharable. However, the agents can respond to $F$ by generating an audio signal through their audio decoder, transforming the internal visual signal:

$$F_i = D_i f_i = D_i E_i F, \quad i = 1, \ldots, N$$

Subsequently, each agent receives two types of secondary signals: one from itself:

$$f_{ii} = E_i F_i = E_i D_i E_i F, \quad i = 1, \ldots, N$$

and others from the rest of the agents:

$$f_{ik} = E_i F_k = E_i D_k E_k F, \quad i, k = 1, \ldots, N, \quad i \neq k$$

Through the "parrot effect," agents adjust their decoder parameters to align their own secondary signals with those from others, aiming for:

$$E_i D_i E_i F \approx E_i D_k E_k F, \quad i \neq k$$

Once this iterative process converges, the distinctions between the signals diminish, leading to a uniform perception by all agents. If the number of agents is sufficiently large, the average

$$\text{Obj}(F) = \text{Avg}_{\{ik\}}(E_i D_k E_k F)$$

serves as an objective representation of the original signal $F$, illustrating how collective interaction among parroting autoencoders can lead to a shared, objective understanding of an observed phenomenon.